**Topological delocalization against Anderson localization observed in Bi-doped PbSb$_2$Te$_4$ disordered topological insulator**

Yuya Hattori[1, 2, a)], Koichiro Yaji[1,3], Keisuke Sagisaka[1], Shunsuke Yoshizawa[1], Shunsuke Tsuda[1], Yuki Tokumoto[2], Taichi Terashima[4], Shinya Uji[4], and Keiichi Edagawa[2]

[1]*Center for Basic Research on Materials, National Institute for Materials Science, 1-2-1 Sengen, Tsukuba, Ibaraki 305-0047, Japan*

[2]*Institute of Industrial Science, The University of Tokyo, Komaba, Meguro-ku, Tokyo 153-8505, Japan*

[3] *Unprecedented-scale Data Analytics Center, Tohoku University, 468-1 Aoba, Aramaki-Aza, Aoba-ku, Sendai 980-8578, Japan.*

[4]*Research Center for Materials Nanoarchitectonics (MANA), National Institute for Materials Science, Sakura, Tsukuba, Ibaraki 305-0003, Japan*

**Abstract**

We experimentally investigate the tendency of localization in the bulk and the topological surface states in topological insulators Pb(Bi$_{1-x}$Sb$_x$)$_2$Te$_4$ ($x = 0.793 \sim 0.818$) through detailed transport measurements. The bulk electronic states in the range $0.793 \leq x < 0.818$ are situated on the insulator side of the Anderson transition, as indicated by $k_F l$ values (where $k_F$ is the Fermi wavenumber and $l$ is the mean free path) falling below the Ioffe-Regel criterion (i.e., $k_F l < 1$). In contrast, the topological surface states retain high mobility and even exhibit quantum oscillations, demonstrating their resilient nature against strong disorder. These findings highlight the delocalized nature of the topological surface states despite Anderson localization of the bulk electronic states.

[a])Corresponding author's E-mail: hattori.yuya@nims.go.jp

The concept of Anderson localization [1–4] has greatly advanced our understanding of the quantum properties of disordered materials. The Anderson transition arises from the interference of electron wave propagation and describes the metal-insulator transition driven by disorder [1,2]. The tendency of localization is generally more pronounced in low dimensional systems, and the scaling theory of Abrahams et al. [2] showed that electron systems with dimension $d \leq 2$ are always localized. The theory was later extended by Hikami, Larkin, and Nagaoka [5] to include magnetic fields and spin-orbit coupling, showing that localization can be weakened to exhibit weak anti-localization (WAL). The theory also provides a useful formula for estimating key transport parameters from magnetoresistance (MR) data.

The surface state of a three-dimensional topological insulator (hereafter, topological surface states) is a notable example of a two-dimensional (2D) system that is expected to be immune to Anderson localization [6–8]. The topological surface state has lifted spin degeneracy and is approximated by massless Dirac electrons [9]. It has been predicted that Anderson localization does not occur in topological surface states because of the $\pi$ Berry phase of Dirac electrons [6–8]. Since the theoretical proposal of three-dimensional topological insulators [10], the topological surface states and their helical spin texture have been observed by angle-resolved photoemission spectroscopy (ARPES) [11–15]. The suppression of backscattering in the topological surface states has also been discussed on the electron interference patterns observed by scanning tunneling microscopy (STM) [16–20]. The transport properties of surface states have also been studied extensively [21–25], and WAL is often observed [21,22,24,25]. Although this phenomenon is indicative of the suppression of localization, it generally occurs in systems with strong spin-orbit coupling [5,26] and does not mean the absence of Anderson localization in the presence of strong disorder. It is therefore difficult to experimentally demonstrate the delocalized nature by focusing only on the topological surface states.

The most direct way to verify the delocalized nature of topological surface states is to contrast the response of conventional electronic states and topological surface states to the same degree of strong disorder. Such an experiment can be performed by measuring the transport properties of both the topological surface and the bulk states, while introducing crystalline disorder strong enough to induce Anderson localization in the bulk states. Although ARPES studies have investigated the existence of the topological surface states under strong disorder [27–29], a detailed characterization of the transport properties is lacking.

Here, we experimentally investigate the delocalized nature of topological surface states against strong disorder in $Pb(Bi_{1-x}Sb_x)_2Te_4$ topological insulators. These compounds are strong topological insulators for all $x$ values between 0 and 1 [12,30]. For intermediate $x$ values, Bi and Sb serve as atomic-scale disorder while preserving the topological index of the material. Through detailed transport measurements, we demonstrate that the bulk transport property exhibits an insulating behavior attributed to the Anderson transition in samples with $0.793 \leq x < 0.818$. This is supported by $k_F l$ values below the Ioffe-Regel criterion ($k_F l < 1$), where $k_F$ is the Fermi wave vector and $l$ is the mean free path. In contrast, transport measurements on nanoflakes show that the topological surface states maintain their delocalized nature in the presence of strong disorder, and even exhibit quantum oscillations, which is a hallmark of high mobility. The coexistence of the bulk states in the Anderson localization regime and topological surface states with high mobility verifies topological delocalization of the topological surface states.

The single crystals of Pb(Bi$_{1-x}$Sb$_x$)$_2$Te$_4$ were grown by the vertical Bridgman method [31]. Phase identification and composition analyses were performed by powder x-ray diffraction (XRD; RIGAKU RINT2500V) and electron probe micro analyzer (EPMA; JEOL JXA-8800RL). The Sb concentration $x$ represents the value measured by EPMA (not nominal composition $x$). Resistivity and Hall measurements were carried out by the standard four-contact resistance measurements from $T = 300$ K to $2$ K in magnetic fields up to $B = 9.0$ T in a commercial physical property measurement system (PPMS, Quantum Design). The Ohmic electrical contacts were made by a room-temperature cured silver paste for thick samples ($t \sim 200$ μm). Additionally, resistivity measurements were performed on several nanoflakes with thicknesses ranging from $t = 400$ nm to $t = 80$ nm, which were exfoliated by the Scotch-tape method on SiO$_2$/Si substrates. Sample cutting and Pt deposition on the nanoflakes were performed using a focused ion beam (FEI, Helios Nano Lab 600i). The ARPES measurements were carried out using the laser-ARPES machine at the Institute for Solid State Physics, the University of Tokyo [32]. The photon energy of the laser was 6.994 eV. The sample temperature was set to 30 K during the ARPES measurements. The STM measurement was performed using an ultrahigh vacuum cryogenic STM (USM-1300, Unisoku co., ltd.) with a mechanically sharpened Pt-Ir tip. In the ARPES and STM measurements, clean (0001) surfaces were prepared by cleaving single crystals of Pb(Bi$_{1-x}$Sb$_x$)$_2$Te$_4$ ($x = 0.80 \pm$ 0.015). The density functional theory calculation of the band structure of PbSb$_2$Te$_4$ was performed by using the OpenMX code [33] with the Perdew-Burke-Ernzerhof generalized gradient approximation [34]. The crystal structure was modeled by a repeated slab of three unit cells and a vacuum layer of 4.5 nm thickness. Details of the computational conditions can be found in the previous report [35].

First, we investigate the bulk electronic properties using thick samples of Pb(Bi$_{1-x}$Sb$_x$)$_2$Te$_4$. Figure 1(a) displays the temperature dependence of the resistivity $\rho$ for Pb(Bi$_{1-x}$Sb$_x$)$_2$Te$_4$ thick samples ($t \sim 200$ μm) with Sb concentrations $x = 0.793$ [31], 0.803 [31], and 0.818. The Sb concentration was precisely controlled by utilizing the concentration gradient along the single crystal growth direction (see Fig. S1 in [35] for details). The data presented in Fig. 1(a) demonstrate that even a small variation of $x =$ [Sb]/([Sb] + [Bi]) in Pb(Bi$_{1-x}$Sb$_x$)$_2$Te$_4$ around $x = 0.8$ induces a pronounced change in the resistivity and its temperature dependence. In the thick samples ($t \sim 200$ μm), the surface conductance is negligible ($\sim 0.1\,\%$, see the Supplementary Material (SM)), indicating that the total conductance is predominantly determined by the bulk states. Samples with $x = 0.793$ and $x = 0.803$ exhibit a bulk insulating behavior [31], characterized by an increase in the resistivity as temperature decreases. In contrast, the $x = 0.818$ sample shows a basically metallic temperature dependence with reduced resistivity values. In this range of Sb concentrations, the electrical conduction is p-type, and the carrier concentrations $n$ determined by Hall measurements range from $\sim 10^{18}$ cm$^{-3}$ to $\sim 10^{19}$ cm$^{-3}$ as shown in Table I. These values are relatively high compared to those of highly bulk-insulating topological insulators such as Sn$_{0.02}$Bi$_{1.08}$Sb$_{0.9}$Te$_2$S ($n < 3 \times 10^{14}$ cm$^{-3}$), where the Fermi energy $E_F$ is situated in the band gap [36]. For our samples, the carrier concentrations correspond to the $E_F$ values of roughly 7 to 30 meV below the edge of the valence band when we assume an effective mass $m^* = 0.5\,m_e$ [37,38]. This indicates that the Fermi energies of our samples cross the top of the valence band.

The finite density of states (DOS) at $E_F$ is also supported by ARPES measurements. Figures 1(b) and 1(c) show the ARPES energy dispersion curves of Pb(Bi$_{1-x}$Sb$_x$)$_2$Te$_4$ ($x = 0.70 \pm 0.015$, and $x = 0.80 \pm 0.015$) along the $\overline{\Gamma}\overline{M}$ direction [39]. The shape of the bands does not change significantly in this composition range, suggesting that the increase in $x$ only induces a decrease of $E_F$ in a rigid-band manner, as previously reported in [13]. In the $x = 0.70$ sample, the topological surface states are identified as an X-shaped bands with the crossing point at a binding energy of $E_B \sim 0.2$ eV (indicated by white dashed lines in Fig. 1(b)). These dispersion curves qualitatively agree with the band calculation result of PbSb$_2$Te$_4$ shown in Fig. 1(d) [35]. While $E_F$ in the $x = 0.70$ sample crosses the bulk conduction band, in the $x = 0.80$ sample, it intersects the bulk valence band around the $\overline{\Gamma}$ point, indicating a finite DOS at $E_F$.

The insulating behavior of the $x = 0.793$ and $0.803$ samples despite the finite density of states at $E_F$ suggests that they are in the Anderson localization regime. This is further confirmed by examining the $x$ dependence of the mobility $\mu$ and $k_F l$ values shown in Fig. 1(e). Here, the mobility $\mu$ was calculated by $\mu = 1/(en\rho)$, where $e$ is the elementary charge. The values of $k_F l$ serve as a quantitative indicator of the degree of disorder, and $l$ of the bulk states is estimated using $l = \frac{\mu m^* v_F}{e} = \frac{\mu \hbar k_F}{e}$ assuming a parabolic dispersion near the top of the valence band. The Fermi wavenumber $k_F$ should be evaluated for the bulk states, but the bulk states are not clearly resolved in the ARPES data shown in Fig. 1(c). For this reason, we use $k_F$ of the topological surface states ($k_F = 0.03$ Å$^{-1}$) as an upper limit for that of the bulk states. Consequently, the $k_F l$ values plotted in Fig. 1(e) are slightly overestimated, but they are well below the Ioffe-Regel criterion ($k_F l = 1$) at $x = 0.793$ and $0.803$ with the insulating behavior of the resistivity as shown in Fig. 1(a). In contrast, $k_F l$ is close to unity at $x = 0.818$ with the metallic behavior of the resistivity. This $x$ dependence agrees well with the abrupt change of the mobility and $k_F l$ as a function of the carrier concentration, which is characteristic of the Anderson transition [3]. The Fermi energy in the samples with $0.793 \leq x < 0.818$ is expected to intersect the localized states, whereas it falls within the extended states in the sample with $x = 0.818$.

The bulk transport properties are summarized in Table I, revealing that $l$ for the samples with $x = 0.793$ and $x = 0.803$ is $1.9\,\text{Å}$ and $4.7\,\text{Å}$, respectively. These values are comparable to the lattice constant of $PbSb_2Te_4$ ($a = 4.35\,\text{Å}$ [40]), suggesting the presence of strong scattering centers in the crystal of the $x = 0.793$ and $x = 0.803$ samples. In fact, a previous study on the crystal structure of $PbSb_2Te_4$ reported a significant concentration of antisite defects [40]. Since our samples are a solid solution of $PbSb_2Te_4$ and $PbBi_2Te_4$, it is plausible that they exhibit an even higher degree of the structural disorder. The disordered nature is further corroborated by STM. Figure 1(f) displays the STM topography of sample $x = 0.80$ ($\pm 0.01$) on the (0001) surface, showing atomic-scale contrasts indicative of electronic state inhomogeneity, likely originating from antisite defects. These atomic perturbations can act as scattering centers for conduction electrons, leading to Anderson localization.

Next, we investigated the transport properties of the topological surface states by utilizing nanoflakes to increase the fraction of the surface conductivity in the total conductivity. In the following experiments, the bulk insulating samples ($x = 0.80 \pm 0.01$) are selected for a further characterization. Figure 2(a) shows the temperature dependence of the resistivity for $Pb(Bi_{1-x}Sb_x)_2Te_4$ ($x = 0.80 \pm 0.01$) with different thicknesses $t = 200\,\mu\text{m}$ [31], 400 nm, and 80 nm. The $200\,\mu\text{m}$ sample behaves as an insulator, while the $80\,\text{nm}$ sample conducts as metal, as evidenced by a decrease in resistivity with decreasing temperature. The $400\,\text{nm}$ sample falls into an intermediate regime between the insulating and metallic samples. The resistivity at $T = 2\,\text{K}$ of the 200 μm, 400 nm, and 80 nm samples is 163 mΩcm, 4.8 mΩcm, and 1.6 mΩcm respectively. Here, the ratio of the surface conduction to the total conductance is estimated to be ∼0.1 % for the $t = 200$ μm sample, ∼40 % for the $t = 400$ nm sample, and ∼80 % for the $t = 80$ nm sample (see the SM). In thin samples, where surface state conduction dominates over bulk state conduction, the metallic temperature dependence of resistivity can be attributed to the transport properties of the surface states.

To further confirm that the total conductance in nanoflakes is dominated by the topological surface states, we perform MR measurements under different magnetic field directions. Figure 2(b) shows the MR of the 80 nm thick sample at different field angles, together with the MR of thick samples ($t$∼200 μm) [31] for comparison. In the 80 nm sample, the MR at low magnetic fields depends only on the perpendicular component of the magnetic field, $B\cos\theta$, indicating the 2D nature of the MR. As shown in Fig. 2(b), the MR of the nanoflake qualitatively differs from that of the thick samples [31] ($t$∼200 μm), where the conductance is dominated by the bulk states. The peak feature observed for the thick sample is attributed to the transition from WAL to weak localization (WL), which is typical of conventional electron systems with strong spin-orbit coupling [26,41]. In contrast, no such feature is observed in the nanoflake sample, and the magnetic field dependences of the resistivity can be well explained by the Hikami-Larkin-Nagaoka (HLN) theory as discussed later. We believe that the qualitative difference in the MR between the thick sample ($t$∼200 $\mu$m) and the thin sample ($t = 80$ nm) is due to the origin of the MR; that is, strong spin-orbit scattering of the bulk states for the thick samples [31,42], and non-trivial Berry phase of the topological surface states for the nanoflakes [9,21]. Therefore, the MR in nanoflakes can be attributed to the transport property of the topological surface states.

The delocalized nature of the topological surface states was more closely examined through an analysis of the MR data using the HLN formula. Figure 2(c) displays the MR of another nanoflake sample ($t = 120$ nm) at different temperatures. At low temperatures, the resistance increases rapidly with increasing $B$, which is a typical behavior of 2D WAL. This behavior fits well with the HLN formula [5] given by:

$$\Delta \sigma = \alpha \frac{e^2}{\pi h} \left[ \Psi \left( \frac{1}{2} + \frac{\hbar}{4 e L_\phi^2 B} \right) - \ln \left( \frac{\hbar}{4 e L_\phi^2 B} \right) \right]. \tag{1}$$

Here, $\Psi$ is the digamma function, and $\alpha = -\frac{1}{2} N$, where $N$ is the number of the independent 2D conduction channels [5,9]. The fitting of the MR data using Eq. (1) is presented in the inset of Fig. 2 (c), demonstrating a good agreement with the experimental values, as indicated by the red curves. As shown in Fig. 2(d), the fitting provides the temperature dependence of the phase coherent length $L_\phi$ and $\alpha$ (inset). In the temperature range of $2$ K $\leq T \leq 20$ K, $L_\phi$ exhibits a power-low behavior $L_\phi \propto T^{-0.64}$, which is in close agreement with the expected scaling $L_\phi \propto T^{-0.5}$ for dephasing by the electron-electron scattering [43]. The fitted parameter of $\alpha$ is close to $\alpha = -1$. This observation implies that both the two conducting channels of the top and bottom topological surface states contribute to the MR, resulting in $N \sim 2$. As illustrated in Fig 2(c), the WAL changes to the normal MR behavior characteristic of metals ($\Delta R_{xx} \propto B^2$) at $T = 40$ K. This transition indicates that the interference of the electron wave propagation is no longer possible at this temperature, and that $L_\phi$ of the topological surface states becomes comparable to $l$. Accordingly, $l$ of the topological surface states is roughly estimated to be $l \sim L_{\phi, T=40 \text{ K}} \sim 30$ nm, which is approximately 70 times longer than that of the bulk states (see Table I). This is surprising since the defect density on the surface should be comparable to or even higher than that in the bulk due to possible surface contamination. These results provide compelling evidence for the delocalized nature of the topological surface states, even in the presence of strong disorder [7,8].

The preservation of high mobility in the topological surface state, despite Anderson localization in the bulk states, is further corroborated by the observation of quantum oscillations. Figure 2(e) shows the resistivity oscillations detected for a nanoflake with thickness $t = 170$ nm. The data were obtained by subtracting a smooth background from the MR signal. Although the signal is relatively weak, the oscillations periodic in $1/B$ are reproducibly observed, and disappear at $T = 5$ K. We ascribe these oscillations to the quantum oscillations of the topological surface states. The oscillation frequency is approximately 45 T, corresponding to $k_F = 0.037$ Å$^{-1}$. The cyclotron mass is estimated to be $m_c = 0.13\, m_0$ as shown in Fig. 2(f). The $k_F$ value agrees with that obtained in the ARPES ($k_{F,ARPES} = 0.03$ Å$^{-1}$) in Fig. 1(c), and the $m_c$ value is also consistent with that estimated from the ARPES data ($m_{c,ARPES} = \hbar k_F / v_F = 0.10 m_0$). Quantum oscillations were also observed in other nanoflake samples with the similar frequencies and cyclotron masses $m_c$ (see Fig. S1 in the SM).

For spintronics applications of topological insulators, a large fraction of surface conductance over bulk conductance is desired [44,45]. For this goal, we propose the use of Anderson localization of the bulk states to suppress the bulk conduction. Historically, the bulk insulation of topological insulators has been achieved by pinning $E_F$ in the band gap, but this strategy requires a precise tuning of the Fermi level at deep impurity levels and its application is limited to a few systems [23,36,46]. In contrast, the Anderson transition of the bulk states can be induced in any materials if $E_F$ is tunable near the edge of the bulk valence/conduction band. The generation of highly polarized spin current would be possible in a wide range of materials using the bulk Anderson transition.

In summary, we have experimentally demonstrated the delocalized nature of topological surface states against strong disorder in a Pb(Bi$_{1-x}$Sb$_x$)$_2$Te$_4$ ($0.793 \leq x < 0.818$) topological insulator. Through detailed transport measurements, we have shown that the bulk electronic structure is in the Anderson localization regime characterized by $k_F l$ values below the Ioffe-Regel criterion. In contrast, the topological surface states retain their delocalized nature against disorder, and even exhibit quantum oscillations. The coexistence of the bulk states in the Anderson localization regime and delocalized topological surface states is a promising condition to achieve highly polarized spin currents for spintronics applications.

## Acknowledgements

YH acknowledges financial support by JSPS KAKENHI (Grant No. JP19J13968 and JP24KJ0227), KS and SY by JP24K01351, KE by JP22H01765. A part of this work was conducted at the University of Tsukuba Nanofabrication Platform in the "Nanotechnology Platform Project" sponsored by MEXT, Japan. MANA was established by World Premier International Research Center Initiative (WPI), MEXT, Japan.

**Table and caption**

Table I: Summary of the bulk transport properties of Pb(Bi$_{1-x}$Sb$_x$)$_2$Te$_4$ samples.

| composition $x$ | $\rho_{2K}$ (m$\Omega$cm) | carrier type | $n$ (cm$^{-3}$) | $\mu$ (cm$^2$/Vs) | $l$ (Å) | $k_F l$ |
|---|---|---|---|---|---|---|
| 0.793 [31] | 67.5 | p-type | $9.8 \times 10^{18}$ | 9.4 | 1.9 | 0.06 |
| 0.798 [31] | 161 | p-type | $3.8 \times 10^{18}$ | 10 | 2.0 | 0.06 |
| 0.803 [31] | 184 | p-type | $1.4 \times 10^{18}$ | 24 | 4.7 | 0.14 |
| 0.818 | 3.79 | p-type | $1.0 \times 10^{19}$ | 164 | 32.4 | 0.97 |

**Figures and captions**

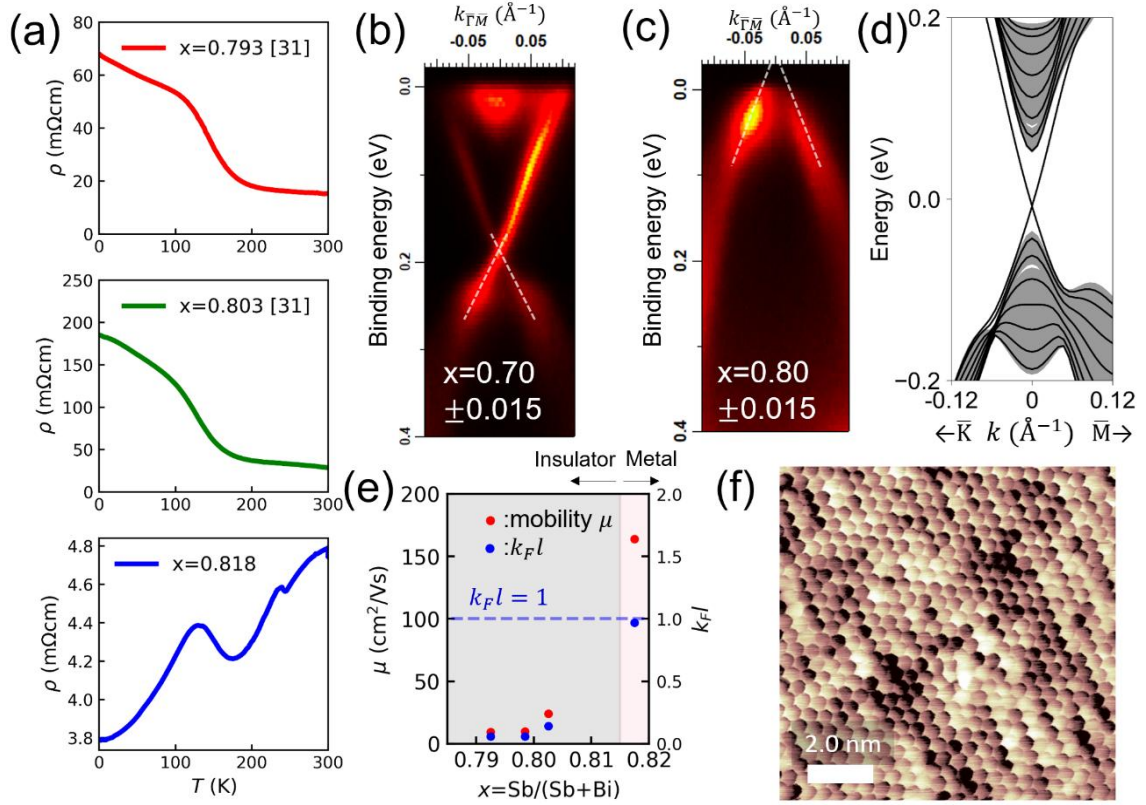

FIG. 1. (a) Temperature dependence of the resistivity for Pb(Bi$_{1-x}$Sb$_x$)$_2$Te$_4$ ($x = 0.793$ [31] , 0.803 [31] , 0.818). (b), (c) ARPES energy dispersion curves along the $\bar{\Gamma}\bar{M}$ direction for Pb(Bi$_{1-x}$Sb$_x$)$_2$Te$_4$ ($x = 0.70 \pm 0.015$ (b) and $0.80 \pm 0.015$ (c)) [39]. (d) Calculated band structure of PbSb$_2$Te$_4$ calculated using a slab model [35]. (e) $x$ dependence of the mobility $\mu$ and $k_F l$. (f) STM topography taken on the (0001) surface of Pb(Bi$_{1-x}$Sb$_x$)$_2$Te$_4$ ($x = 0.80 \pm 0.01$).

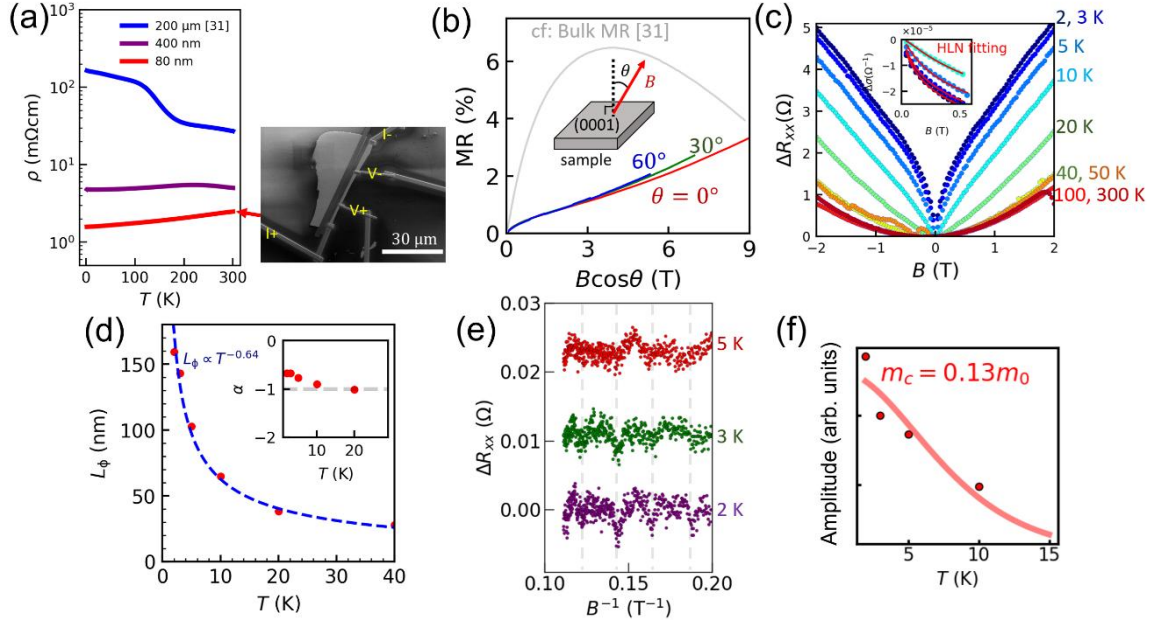

FIG. 2. (a) Temperature dependence of the resistivity for $Pb(Bi_{1-x}Sb_x)_2Te_4$ ($x = 0.80 \pm 0.01$) samples with different thicknesses $t = 200\ \mu m$ [31], 400, and 80 nm. The inset displays a scanning electron microscopy image of the $t = 80\ nm$ sample. (b) MR of the $t = 80\ nm$ sample at different magnetic field angles, compared with the MR of the bulk states [31]. The inset illustrates the definition of a magnetic field angle. (c) MR at different temperatures. The inset presents the fitting results with the HLN formula. (d) Temperature dependence of the fitting parameter $L_\phi$, and the inset shows the temperature dependence of $\alpha$. (e) Quantum oscillations observed in a nanoflake of $Pb(Bi_{1-x}Sb_x)_2Te_4$ ($x = 0.80 \pm 0.01$). (f) Temperature dependence of the oscillation amplitude in (e).

**Supplementary Materials:**

**Topological delocalization against Anderson localization observed in Bi-doped**

**PbSb$_2$Te$_4$ disordered topological insulator**

Yuya Hattori[1, 2, a)], Koichiro Yaji[1,3], Keisuke Sagisaka[1], Shunsuke Yoshizawa[1],Shunsuke Tsuda[1], Yuki Tokumoto[2], Taichi Terashima[4], Shinya Uji[4], and Keiichi Edagawa[2]

[1]*Center for Basic Research on Materials, National Institute for Materials Science,*

*1-2-1 Sengen, Tsukuba, Ibaraki 305-0047, Japan*

[2]*Institute of Industrial Science, The University of Tokyo, Komaba, Meguro-ku, Tokyo 153-8505, Japan*

[3] *Unprecedented-scale Data Analytics Center, Tohoku University, 468-1 Aoba, Aramaki-Aza, Aoba-ku, Sendai 980-8578, Japan.*

[4]*Research Center for Materials Nanoarchitectonics (MANA), National Institute for Materials Science, Sakura, Tsukuba, Ibaraki 305-0003, Japan*

Corresponding authors' E-mail: HATTORI.Yuya@nims.go.jp (Yuya Hattori),

## I. Ratio of surface conduction to total conduction

The conductance of the surface states $G_s$ and bulk conductance $G_b$ can be expressed as:

$$G_s = \sigma_s \times \frac{w}{L} = en_s\mu_s \times \frac{w}{L},$$

$$G_b = \sigma_b \times \frac{wt}{L} = en_b\mu_b \times \frac{wt}{L},$$

where $\sigma_i$, $n_i$, and $\mu_i$ are the conductivity, carrier concentration, and the mobility of a conduction channel $i$. The parameters $w$, $L$, and $t$ are the width, length, and the thickness of the sample for resistivity measurements. Thus, the ratio of the surface conduction to the bulk conduction is:

$$G_s : G_b = n_s\mu_s : n_b\mu_b \times t.$$

The mobility of the topological surface states can be calculated as:

$$\mu_s = \frac{e\tau}{m_c} = \frac{el}{\hbar k_F}.$$

Using the value of the mean free path $l \sim 30$ nm obtained from the WAL analyses and $k_F = 0.037$ Å$^{-1}$ from the SdH oscillations, the mobility of the surface states in Pb(Bi$_{0.21}$Sb$_{0.79}$)$_2$Te$_4$ is estimated to be $\mu_s \sim 1200$ cm$^2$/(Vs). The carrier concentration of the surface states can be calculated from the SdH oscillations, and the value is $n_s = 2.1 \times 10^{12}$ cm$^{-2}$. The mobility and carrier concentration of the bulk states are estimated from the Hall measurements, and the values are $\mu_b = 9.4$ cm$^2$/(Vs) and $n_b = 9.8 \times 10^{18}$ cm$^{-3}$ respectively.

For a sample with a thickness of $t = 200$ μm, the ratio of the surface conductance to the total conductance is $G_s/(G_s + G_b) \sim 0.1\%$. Therefore, the MR of the thick sample ($t = 200$ μm) in Fig. 2(b) stems from the bulk conduction. On the other hand, in the thin sample ($t = 80$ nm), the ratio of the surface conductance to the total conductance is $G_s/(G_s + G_b) \sim 80$ %.

## II. Quantum oscillations in other samples

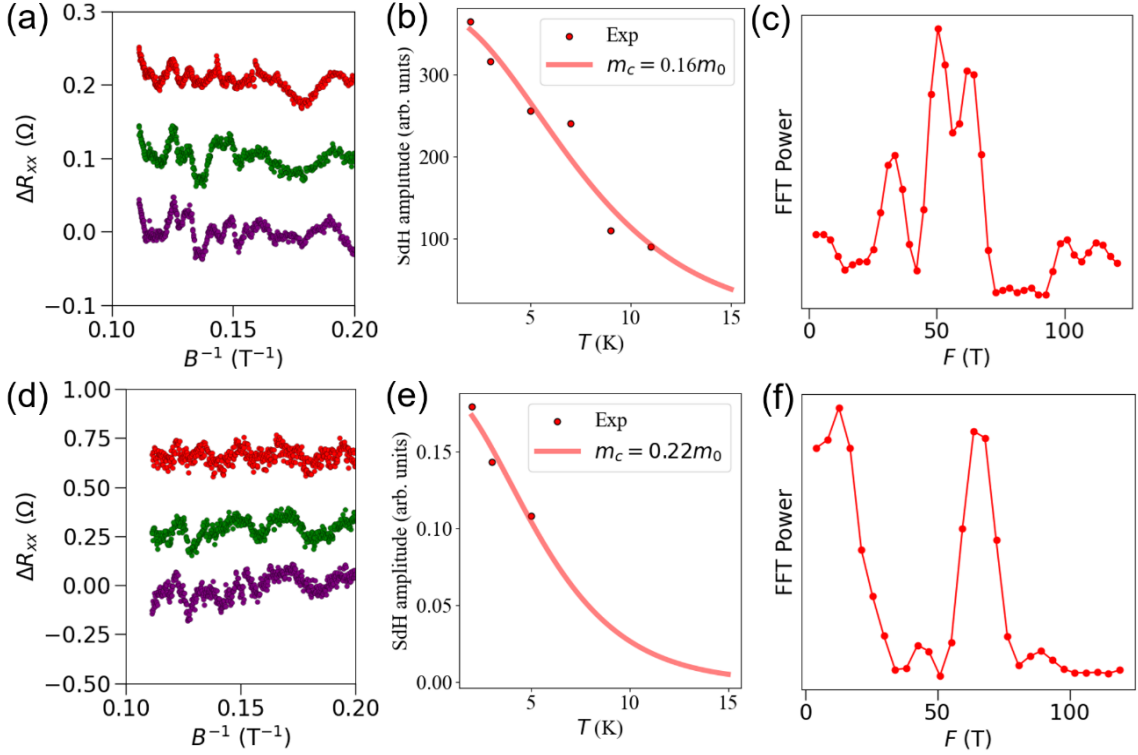

Fig. S1 (a-c) The quantum oscillation (a), and mass fitting (b), and power spectrum (c) using Fast Fourier Transform (FFT) in sample S1. (d-f) The quantum oscillation (d), and mass fitting (e), and power spectrum (f) using FFT in sample S2.

Figure S1 shows the quantum oscillations observed in other nanoflakes of $Pb(Bi_{0.21}Sb_{0.79})_2Te_4$. The frequencies $F$ of the oscillations show slight sample dependence ($F = 57$ T in sample S1, and $F = 66$ T in sample S2). The group velocity of the topological surface states, which is calculated by $v_F = \hbar k_F/m_c$, is $v_F = 1.9$ eVÅ in sample S1 and $v_F = 1.6$ eVÅ in sample S2. These values are roughly consistent with the group velocity observed in the ARPES measurement ($v_F = 2.2$ eVÅ) in the main text.